\documentclass{elsarticle}
\usepackage[utf8]{inputenc}
\usepackage[margin=1.25in]{geometry}
\usepackage{graphicx}
\usepackage{amsmath, amssymb}
\usepackage{float}
\usepackage{placeins}

\title{Speed Control of DC Motor Using Fuzzy PID Controller \tnoteref{t1}}
\tnotetext[t1]{Senior thesis}
\author{Nassim Messaadi
\footnote{Nassim Messaadi : \\ Email address : nassim.messaadi@etu.usthb.dz \\ Address : Cité hospitalière Kouba, bâtiment 1, n°7, 16006 Algiers, Algeria}}

\author{Abdelkader Amroun
\footnote{Abdelkader Amroun  : \\ Email address : abdelkader.amroun@etu.usthb.dz  \\ Address : Villa 13 cité ARMAF, rostomia, Dely brahim, 16047 Algiers, Algeria}}

\author{\footnotesize\\Department of Instrumentation and Automation\\  University of Science and Technology Houari Boumediene }

 \date{August 2021}
 
\begin{document}

\begin{abstract} {
In this project, we designed a DC motor whose speed can be controlled by a PID controller. The proportional, integral and derivative gains (KP, KI, KD) of the PID controller are adjusted according to Fuzzy logic. 

First of all, the fuzzy logic controller is designed according to rules so that the systems is basically robust. There are 25 rules for the auto-tuning of each parameter of the PID controller. 
The FLC (fuzzy logic controller) has two inputs. The first is the motor speed error between the reference (setpoint) and the actual speed. The second is the variation of the speed error (derivative of the speed error).
Secondly the output of the FLC is the parameters of the PID controller which are used to control the speed of the DC motor.

The study shows that both the precise characters of PID controllers and the flexible characters of fuzzy controllers are present in the fuzzy self-tuning PID controller.
The fuzzy auto-tuning approach implemented on a conventional PID structure was able to control the speed of the DC motor. It also improved the dynamic and static response of the system.
The comparison between the conventional response and the fuzzy self-tuning response was performed based on the simulation result obtained by MATLAB/SIMULINK.
 The simulation results show that the designed self-adaptive PID controller achieves good dynamic behavior of the DC motor, perfect speed tracking with short rise and settling times, zero overshoot and steady state error and thus gives better performance compared to the conventional PID controller.
We then model the fuzzy PID using simple code on Arduino IDE and  perform a practical experiment, to confirm our theorical results.}
\end{abstract}


\maketitle

\newpage
\section{Introduction}
{
It is well known that, despite the great number of new methodologies that are continuously developed by researchers in the field of automatic control, the most adopted controllers for industrial processes are still the Proportional-Integral-Derivative (PID) controllers.
\\
This is motivated by the fact that, although they usually do not achieve optimal performance, they can guarantee satisfactory results for a wide range of processes and, due to their simple structure, they often represent the best solution from a quality/price point of view.
\\
In this project, a new methodology for tuning PID controllers for processes with underdamped response is proposed. It allows to determine the proportional, integral and derivative gains through a fuzzy inference system.
\\
The main objective of this methodology is to improve the performance of the system in terms of setpoint tracking, i.e. to decrease the rise time, overshoot and response time.
\\
Fuzzy logic has already been adopted in the improvement of PID control and it seems appropriate to use it also in this context, since the design of the tuning procedure can be done in an intuitive way, reflecting the typical experience of the operator.
}

\section{Modeling and Simulation}
\subsection{Modeling of the DC motor}
{The electrical equivalent of a DC motor is illustrated below. It can be represented by a voltage generator (V) in series with an inductance (L) and a resistor (R) and this in series with an induced voltage (e) of opposite direction to our voltage generator. The induced voltage is generated by the rotation of the electric coil through the fixed flux lines of the permanent magnets. This voltage is called electromotive force.  
}

\FloatBarrier
\begin{figure}[htp]
    \centering
    \includegraphics[width=9cm]{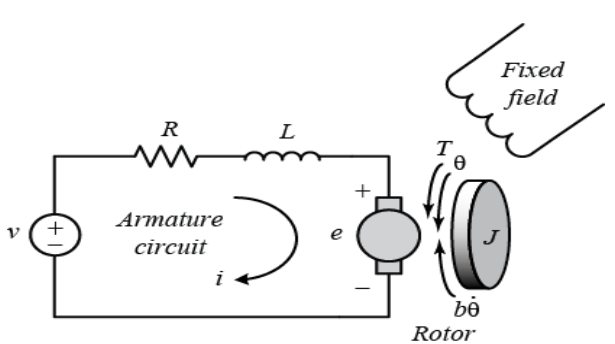}
    \caption{ Representation of a DC motor}
    \label{fig:DCmotor}
\end{figure}
\FloatBarrier

{
In general, the torque generated by a DC motor is proportional to the armature current and the magnetic field strength. Assume that the magnetic field is constant and therefore the motor torque (T) is proportional to the armature current (i) alone by a constant factor (Kt) called the torque constant, as shown in equation (1). This is called an armature-driven motor.}

\begin{equation} \label{1.1} T = K_t \times i \end{equation}
{
The back electromotive force emf (e) is proportional to the angular velocity of the shaft ($\theta$) by a constant factor (Kb) called the electromotive force constant, as shown in Figure 1.}
\begin{equation} \label{1.2} e = K_b \times \theta \end{equation}
{
We derive the following governing equations based on Newton's second law in equation (3) and Kirchhoff's mesh law in equation (4).}
\begin{equation} \label{1.3} J \ddot \theta + b \dot \theta =K_t \times i \end{equation}
\begin{equation} \label{1.4} La \times \frac{di}{dt} + Ra \times i = V - K_b \times \dot \theta \times b \end{equation}
{
(J) is the moment of inertia of the rotor, (b) is the viscous friction constant of the motor, (La) is the electrical inductance, (Ra) is the electrical resistance, and (V) is the voltage source. By applying the Laplace transform, the modeling equations can be expressed in terms of Laplace variables as follows in terms of Laplace variables, as shown in equations (5) and (6).}
\begin{equation} \label{1.5} s(J.s+b)\times \Theta(s) = K_t \times I(s)  \end{equation}
\begin{equation} \label{1.6} s(La.s+Ra)\times I(s) = V(s)- K_b \times s.\Theta(s)  \end{equation}
{
The open-loop transfer function below was obtained by eliminating I(s) in equations (7), where the rotational speed is considered as the output and the armature voltage as the input.}
\begin{equation} \label{1.7} G(s)=\frac{\Theta(s)}{V(s)}= \frac{K_t}{(j.s+b)(La.s+Ra)+K_t\times K_b} \; \; \;  [\frac{rad/sec}{V}]  \end{equation}
 \\
 \FloatBarrier
\begin{table}[ht]
\begin{center}
 \begin{tabular}{|c|c|} 
 \hline
 Parameters & Value \\ [0.5ex] 
 \hline\hline
  Torque constant (Kt) & 0.5 N.m/A \\ 
 \hline
  Electromotive force constant (Kb) & 1.25 V/rad/s \\
 \hline
  Electrical resistance (Ra) & 5$\Omega$ \\
 \hline
  Viscous friction constant of the motor (b) & 0.008 N.m/rad/s \\
 \hline
  Electrical inductance (La) & 0.2 H \\
 \hline
  Moment of inertia of the rotor (J) & 0.1 kg.m²  \\ [1ex] 
 \hline
\end{tabular}
\caption{ Parameters of the DC motor}
\label{tab:Parameters}
\end{center}
\end{table}
\FloatBarrier
{
Using the parameters of our engine which are in the table above in equation (7), we obtain the following equation:}

\begin{equation} \label{1.8} G(s)=\frac{\Theta(s)}{V(s)}= \frac{0.5}{(0.1s+0.008)(0.2s+5)+ 0.5 \times 1.25} \; \; \;  [\frac{rad/sec}{V}]  \end{equation}
{
We can then translate the previous equation by the Simulink diagram shown below : \\}
\FloatBarrier
\begin{figure}[htp]
    \centering
    \includegraphics[width=11cm]{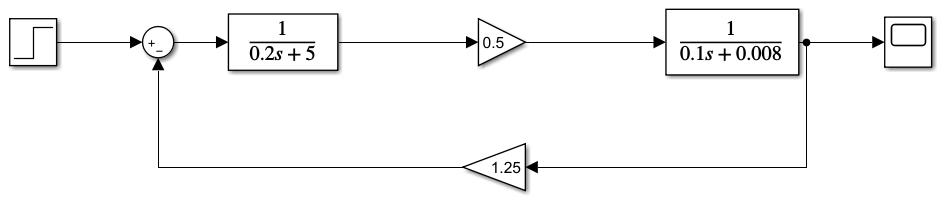}
    \caption{\footnotesize Representation of a DC motor}
    \label{fig:DCmotor parameters}
\end{figure}
\FloatBarrier

\newpage

\subsection{Conventional controller modeling }
\subsubsection{PID controller}
{
The function of the PID controller is mainly to adjust an appropriate proportional gain (KP), integral gain (KI) and differential gain (KD) in order to achieve optimal control performance. \\
There are several methods for adjusting the PID parameters, in our case we will use the "Manual Tuning" method.\\
The manual Tuning of the PID controller is done by setting the reset time to its maximum value and the rate to zero and increasing the gain until the loop oscillates at a constant amplitude.\\
First, we set the ki and kd values to zero. We increase kp until the loop output oscillates, then kp should be set to about half that value for a "quarter-amplitude decay" type response. Then ki is increased until any offset is corrected within a sufficient time for the process. However, a value that is too high will lead to instability. \\
Finally, Kd is increased, if necessary, until the loop is fast enough to reach its set point after a load disturbance. However, increasing Kd too much will result in excessive response and overshoot. Fast PID loop tuning usually results in a small overshoot to reach the setpoint faster; however, some systems cannot accept overshoot, in which case an over-damped closed-loop system is required, which will require tuning well below half the setting that was causing the oscillation.\\ 
The values obtained are: Kp = 30, ki=12 ,kd=1.}

\FloatBarrier
\begin{figure}[htp]
    \centering
    \includegraphics[width=10cm]{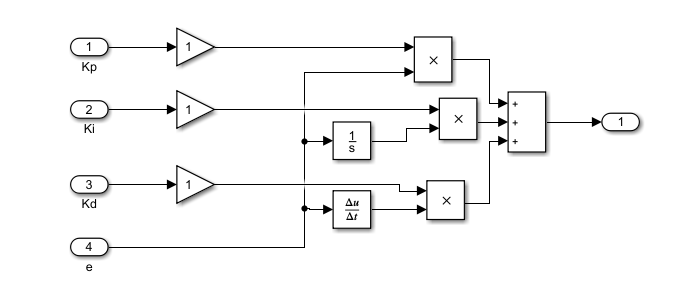}
    \caption{ Simulink diagram of the PID controller}
    \label{fig:PID interne}
\end{figure}
\FloatBarrier

{
We put this controller in series with our motor, we then obtain the Simulink diagram below:}
\FloatBarrier
\begin{figure}[htp]
    \centering
    \includegraphics[width=18cm]{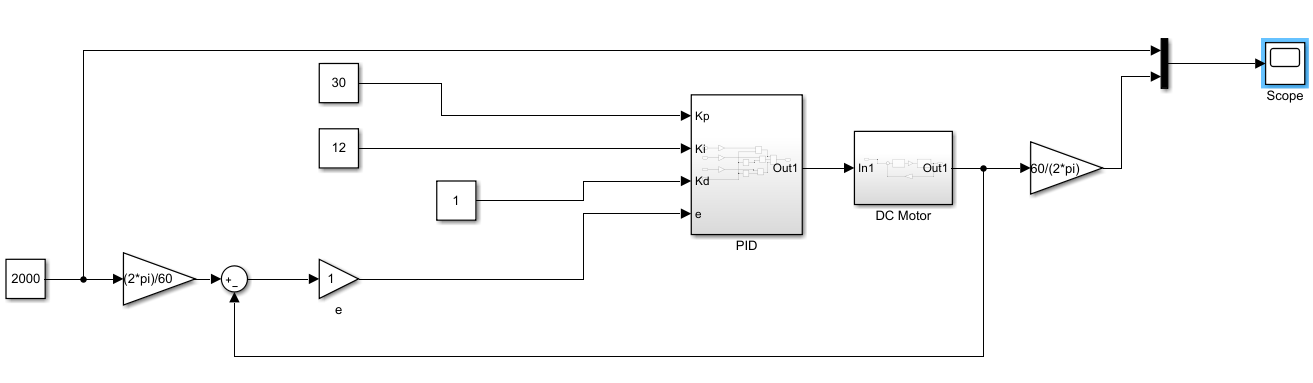}
    \caption{ Simulink diagram of the DC motor in series with a PID controller}
    \label{fig:PID serie}
\end{figure}
\FloatBarrier

\newpage

\subsubsection{Fuzzy PID controller}

{
Fuzzy logic is expressed using human language. Based on fuzzy logic, a fuzzy controller converts a linguistic control strategy into a control strategy, and the fuzzy rules are constructed by the experience of an expert or a database.\\
The error (e) and the change of the error (de) of the speed are the variable inputs of the fuzzy logic controller. The variable output of the fuzzy logic controller adjusts the parameters of the PID controller, then the PID controller calculates the control output.
}

\FloatBarrier
\begin{figure}[htp]
    \centering
    \includegraphics[width=12cm]{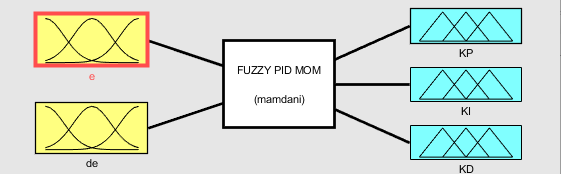}
    \caption{ Inputs and Outputs of the Fuzzy Controller}
    \label{fig:FLC}
\end{figure}
\FloatBarrier
{
All membership functions have an asymmetric form with more concentration near the origin. This allows for greater accuracy in the steady state.
Figures (6 and 7) show the input values (error and error change) that are applied in the fuzzy PID controller.}

\FloatBarrier
\begin{figure}[htp]
    \centering
    \includegraphics[width=8cm]{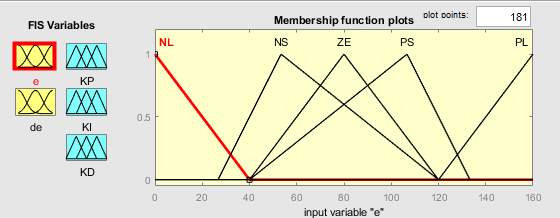}
    \caption{ Membership function of the Error}
    \label{fig:e fuzzy}
    \centering
    \includegraphics[width=8cm]{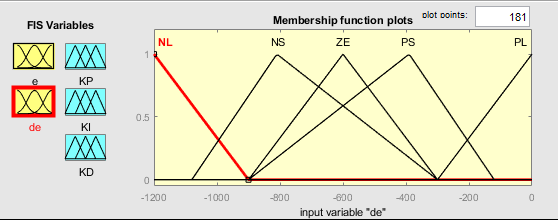}
    \caption{ Membership function of the Error change}
    \label{fig:de fuzzy}
\end{figure}
\FloatBarrier

{\footnotesize \; \\ Figures 8, 9 and 10 show the output values (Kp, ki, kd) that are 
applied to the PID controller.\\ \; }

\newpage

\FloatBarrier
\begin{figure}[htp]
    \centering
    \includegraphics[width=8cm]{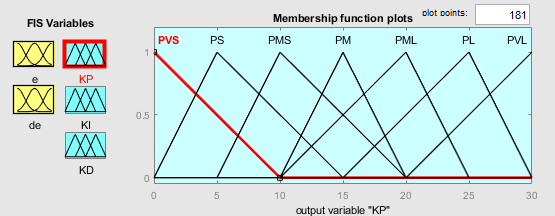}
    \caption{ Membership function of the gain Kp}
    \label{fig:kp fuzzy}
    \centering
    \includegraphics[width=8cm]{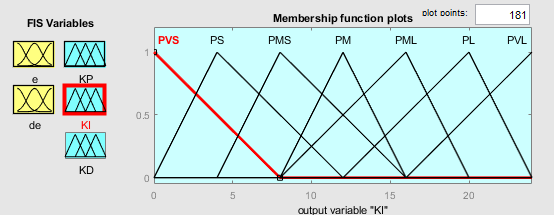}
    \caption{ Membership function of the gain Ki}
    \label{fig:ki fuzzy}
\end{figure}
\FloatBarrier

\FloatBarrier
\begin{figure}[htp]
    \centering
    \includegraphics[width=8cm]{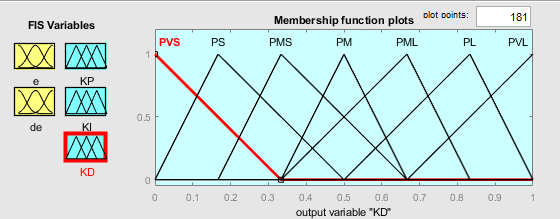}
    \caption{ Membership function of the gain Kd}
    \label{fig:kd fuzzy}
\end{figure}
\FloatBarrier

\FloatBarrier
\begin{table}[H]
\begin{center}
 \begin{tabular}{|c|c|c|c|c|c|} 
 \hline
 e/de & NL & NS & ZE & PS & PL\\ [0.5ex] 
 \hline
  NL & PVL & PVL & PVL & PVL & PVL\\
 \hline
  NS & PML & PML & PML & PL & PVL\\
 \hline
ZE  & PVS & PVS & PS & PMS & PMS \\
 \hline
PS  & PML & PML & PML & PL & PVL\\
 \hline
PL  & PVL & PVL & PVL & PVL & PVL\\ [1ex] 
 \hline 
\end{tabular}
\caption{ The rules applied in the PID controller for Kp}
\label{tab:Kp}
\end{center}
\end{table}
\FloatBarrier

\FloatBarrier
\begin{table}[H]
\begin{center}
 \begin{tabular}{|c|c|c|c|c|c|} 
 \hline
 e/de & NL & NS & ZE & PS & PL\\ [0.5ex] 
 \hline
  NL & PM & PM & PM & PM & PM\\
 \hline
  NS & PMS & PMS & PMS & PMS & PMS\\
 \hline
ZE  & PS & PS & PVS & PS & PS \\
 \hline
PS  & PMS & PMS & PMS & PMS & PMS\\
 \hline
PL  & PM & PM & PM & PM & PM\\ [1ex] 
 \hline 
\end{tabular}
\caption{ The rules applied in the PID controller for Ki}
\label{tab:Ki}
\end{center}
\end{table}
\FloatBarrier

\FloatBarrier
\begin{table}[H]
\begin{center}
 \begin{tabular}{|c|c|c|c|c|c|} 
 \hline
 e/de & NL & NS & ZE & PS & PL\\ [0.5ex] 
 \hline
  NL & PVS & PMS & PM & PL & PVL\\
 \hline
  NS & PMS & PML & PL & PVL & PVL\\
 \hline
ZE  & PM & PL & PL & PVL & PVL \\
 \hline
PS  & PML & PVL & PVL & PVL & PVL\\
 \hline
PL & PVL & PVL & PVL & PVL & PVL\\ [1ex] 
 \hline 
\end{tabular}
\caption{ The rules applied in the PID controller for Kd}
\label{tab:Kd}
\end{center}
\end{table}
\FloatBarrier


{\footnotesize The Simulink diagram of our MCC in series with the fuzzy PID controller is shown in the following figure:}
\FloatBarrier
\begin{figure}[htp]
    \centering
    \includegraphics[width=17cm]{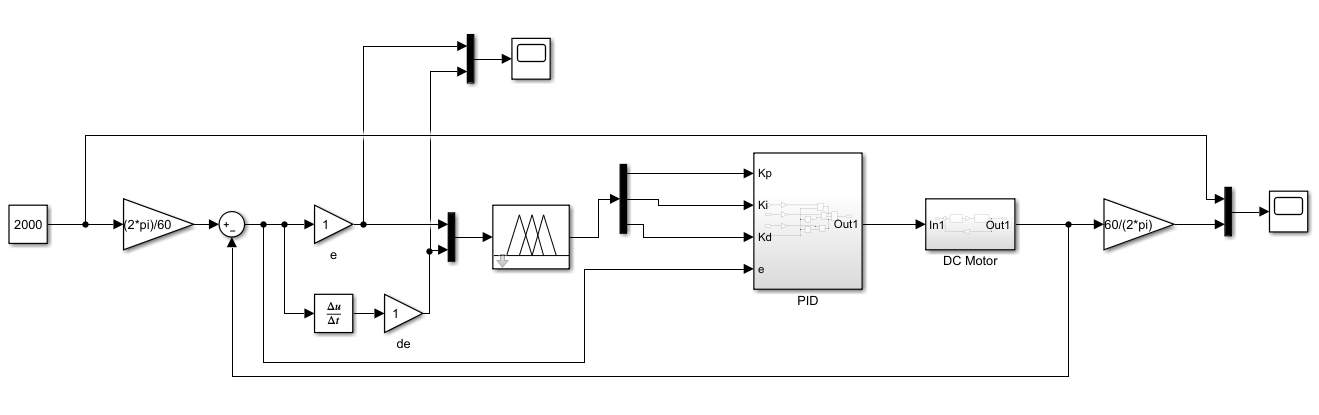}
    \caption{ Simulink diagram of the DC motor in series with a Fuzzy PID Controller}
    \label{fig:fuzzy PID simu}
\end{figure}
\FloatBarrier

{\footnotesize For example, the rules applied in the PID Fuzzy as shown in Figure 12, if the error 'e' is equal to zero (NL) and the variation of the error 'de' is equal to zero (PL), then Kp equal to 30 (PVL), Ki equal to 12 (PM) and kd equal to 1 (PVL).}
\FloatBarrier
\begin{figure}[htp]
    \centering
    \includegraphics[width=13cm]{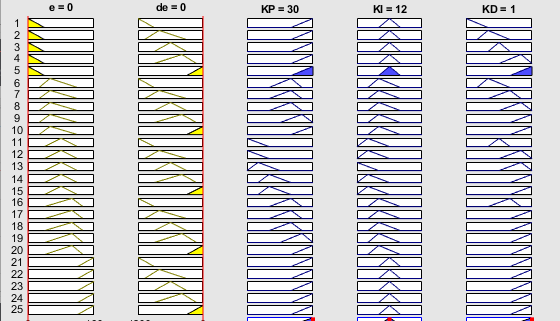}
    \caption{ The Fuzzy PID rule viewer}
    \label{fig:Rules viewer}
\end{figure}
\FloatBarrier

\newpage

\subsection{Simulation}
\subsubsection{System without controller}
{\footnotesize Before performing the control, we must first see the motor output in rpm. In this case (figure 13), the step response does not reach the desired value, the rise time is too high and the steady state error is very high.}
\FloatBarrier
\begin{figure}[htp]
    \centering
    \includegraphics[width=11cm]{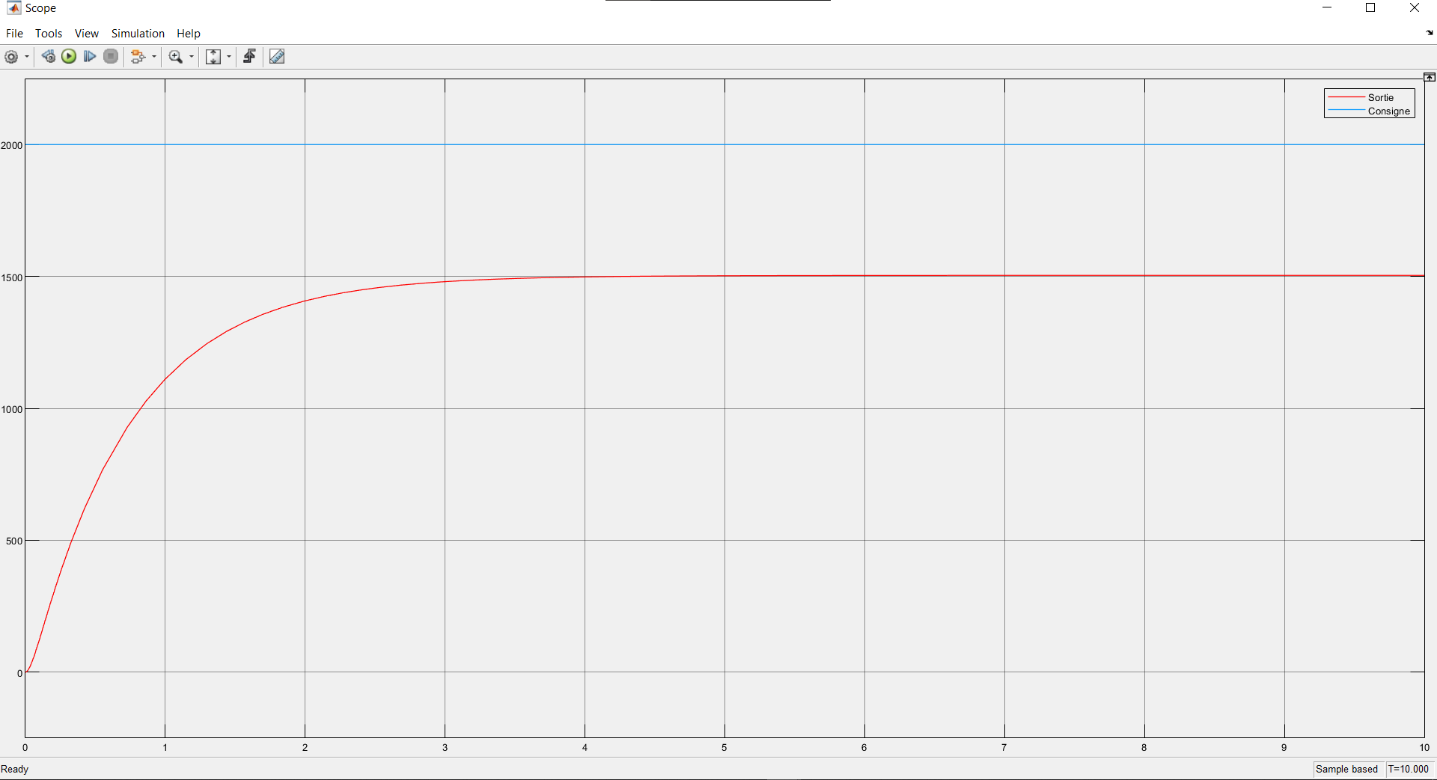}
    \caption{ DC motor output without controller for a setpoint of 2000rpm}
    \label{fig:Dc plot}
\end{figure}
\FloatBarrier

\subsubsection{System using a PID controller}
{\footnotesize In this case (figure 14), the step response reaches the desired value, very small rise time, appropriate settling time and the steady state error is very low (insignificant).}
\FloatBarrier
\begin{figure}[htp]
    \centering
    \includegraphics[width=12cm]{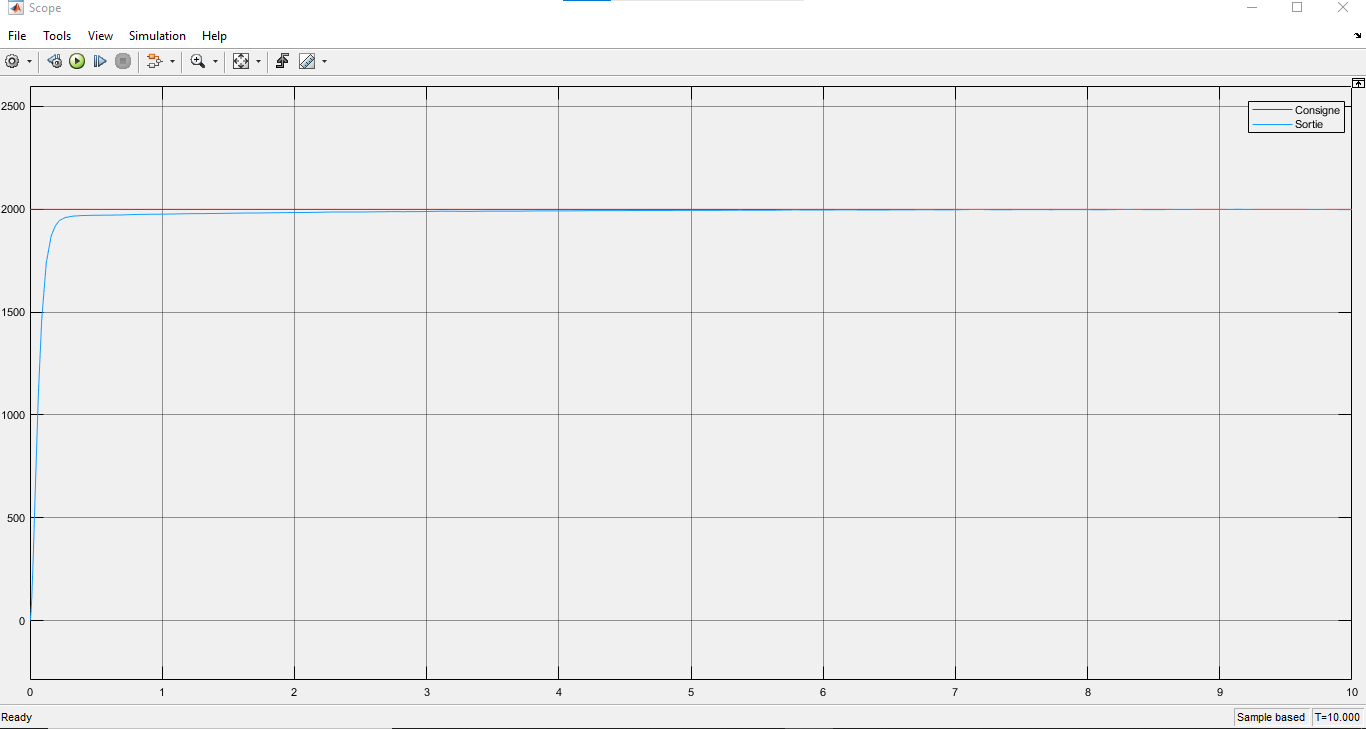}
    \caption{ DC motor output with a PID controller in series}
    \label{fig:PID plot}
\end{figure}
\FloatBarrier

\newpage

\subsubsection{System using a fuzzy PID controller}
{\footnotesize In this case (figure 15), Fuzzy Logic is used for setting the PID parameters. The response reaches the desired value, no steady state error, zero overshoot, the settling and rise times are minimal.}
\FloatBarrier
\begin{figure}[htp]
    \centering
    \includegraphics[width=12cm]{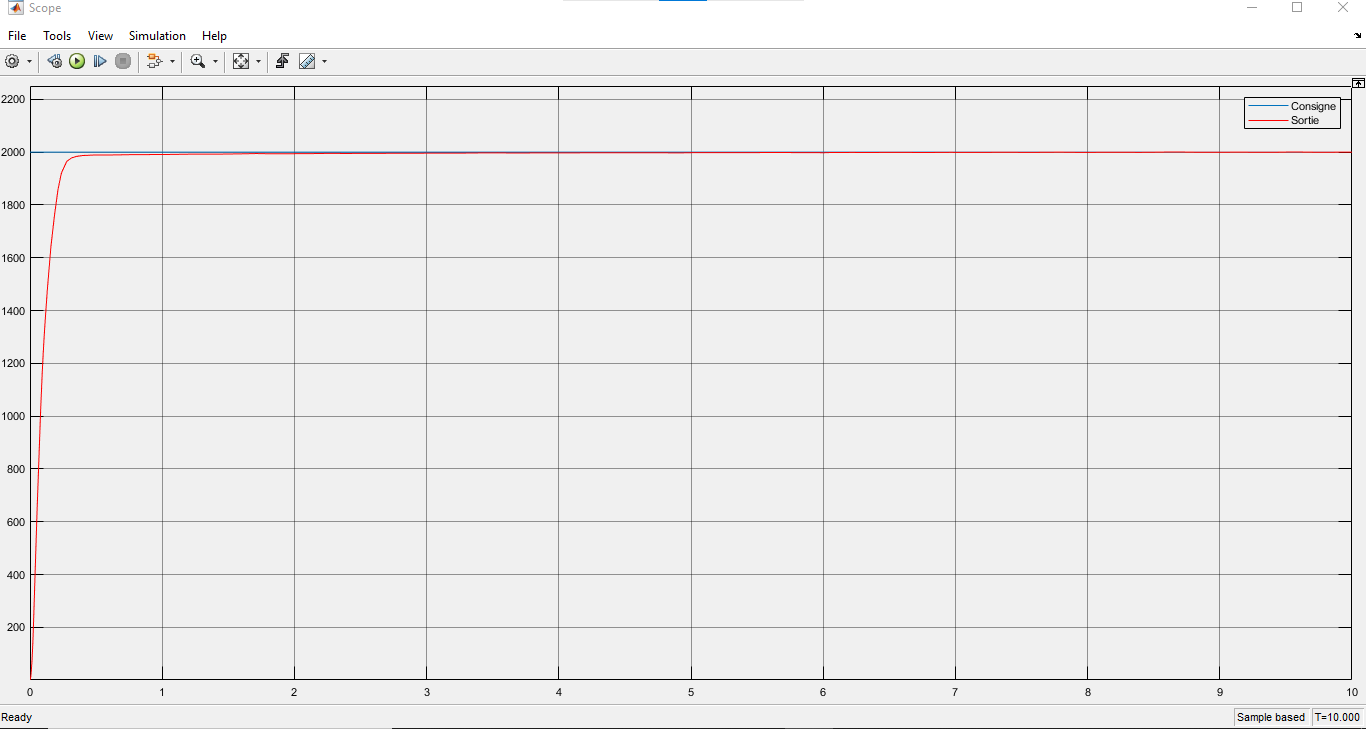}
    \caption{ DC output with Fuzzy PID controller}
    \label{fig:Fuzzy PID plot}
\end{figure}
\FloatBarrier

{\footnotesize Below (figure 16) we find the graphs of the error (in red) and the derivative of the error (in blue), we can see that the error cancels out very quickly, we can also see that the derivative of the error reaches a very large negative value from the first times which is considerably large but this is compensated very quickly.}
\FloatBarrier
\begin{figure}[htp]
    \centering
    \includegraphics[width=12cm]{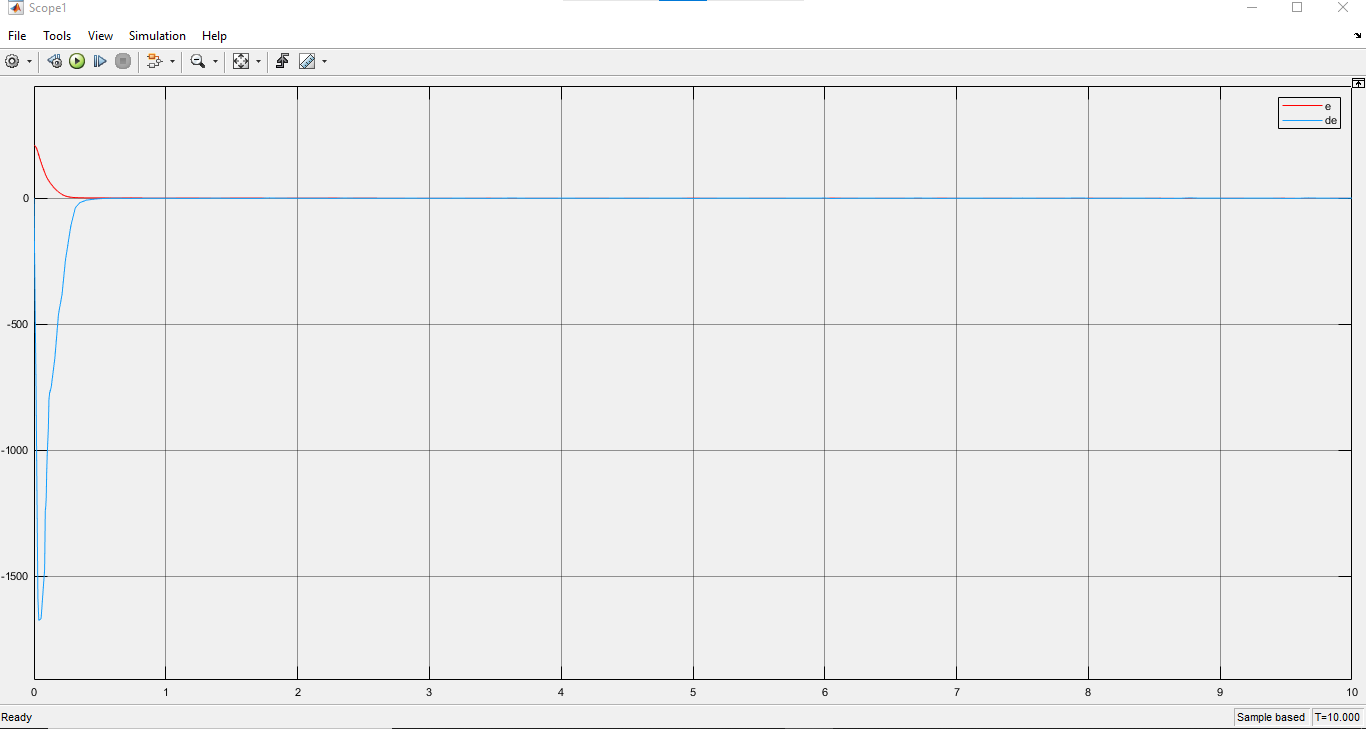}
    \caption{ The error and the derivative of the error of the DC motor in series with a Fuzzy PID controller}
    \label{fig:Fuzzy PID plot}
\end{figure}
\FloatBarrier

\newpage

\subsection{Conclusion}
{ From the table (5) we notice that : \\
- None of the three setups show overshoot.\\
- The rise time of the DC motor without controller is considerably larger than with controller and that the PID controller offers better performance than the fuzzy PID controller in terms of rise time. \\
- The steady state error without controller is very large, the simple PID controller offers negligible error and the fuzzy PID offers no error.\\
- The response time without controller is quite large, the latter improves in the case of the simple PID but remains relatively slow compared to the fuzzy PID which is 4 times faster than the PID and 10 times faster than the DC motor without controller.}
\FloatBarrier
\begin{table}[H]
\begin{center}
 \begin{tabular}{|c|c|c|c|c|c|} 
 \hline
 Parameters & DC motor whitout controller & PID & Fuzzy PID\\ [0.5ex] 
 \hline\hline
Overshoot(\%)  & 0 & 0 & 0\\ 
 \hline
Rise time (10-90\%) (s)  & 1.556 s & 55.791 ms & 170.761 ms\\ 
 \hline
Steady state error  & 500 &1.5  & 0\\ 
 \hline
Settling time (2\%) (s)  & 2.759 & 1.19 & 0.274\\  [1ex] 
 \hline 
\end{tabular}
\caption{ Comparison between the step responses of each system}
\label{tab:Results simu}
\end{center}
\end{table}
\FloatBarrier

\FloatBarrier
\begin{figure}[htp]
    \centering
    \includegraphics[width=12cm]{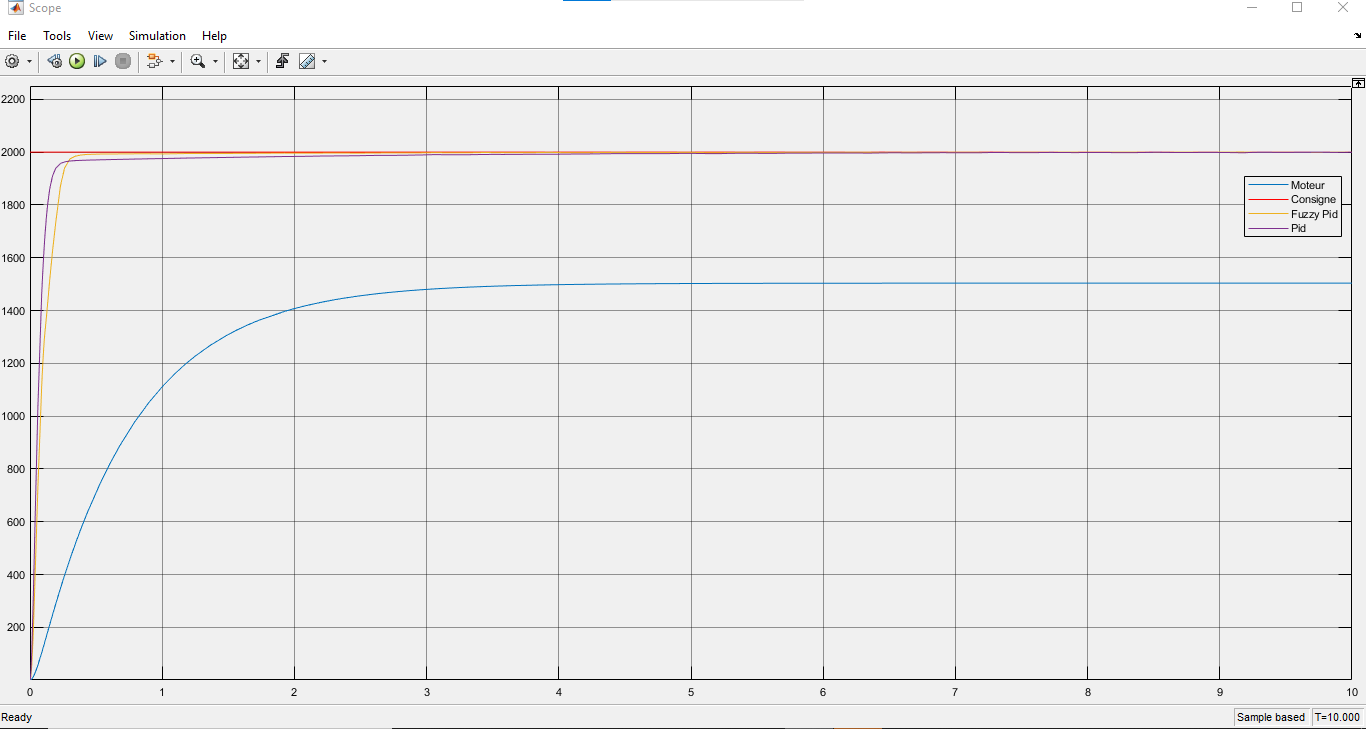}
    \caption{ Comparison between the step responses of each system}
    \label{fig:Fuzzy PID plot}
\end{figure}
\FloatBarrier
{ From table (5) and figure (17) we conclude that:\\
- DC motor alone offers unreliable performance.\\
- The classical PID is faster in rise time than the fuzzy PID but it takes much longer to reach its final value.\\
- The fuzzy PID has the best compromise between rise time and response time, it also offers a very accurate output.
}

\newpage

\section{Practical experiment }
\subsection{Material used}
\FloatBarrier
\begin{table}[H]
\begin{center}
 \begin{tabular}{c|c} 

 - Arduino UNO & - 4 pins fan\\ [0.5ex] 

 - Driver L293D & - Electronic relay module (5V)\\ 

 - Resistance 1K $\Omega$ & - Capacitor 70µF \\ 

- LCD display unit (16x2)  & - Breadboard \\  [1ex] 

\end{tabular}
\caption{ List of material}
\label{tab:material}
\end{center}
\end{table}
\FloatBarrier

\subsection{Diagram of the installation }
{\footnotesize The diagram of the realization is shown in figure (18) below:}

\FloatBarrier
\begin{figure}[htp]
    \centering
    \includegraphics[width=12cm]{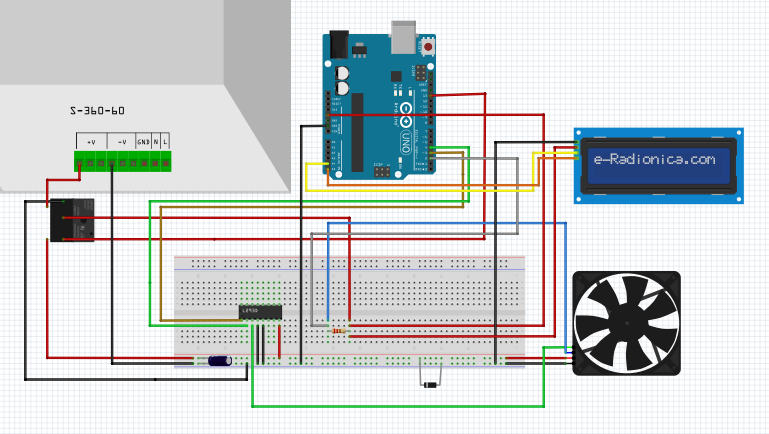}
    \caption{ Installation}
    \label{fig:Installation}
\end{figure}
\FloatBarrier

{ Our system is controlled with an Arduino UNO which is connected to the computer to obtain the information of the system via the Arduino IDE interface, as we can see our system is powered in 12V, in order to be able to control the power supply we use an electronic relay module between the power supply and the board of experimentation. To make our Arduino communicate with our motor we use the L293D driver. The capacitor and the diode between the positive and negative terminals are used for filtering and elimination of parasitic signals, the DC motor in this system is included in the 4 pins fan, including a pin for the +12V, a pin for the ground, a pin linked to a Tacheometer (Hall effect sensor, which generates two pulses per revolution) and finally a PWM pin (pulse width modulation) which allows us to control our DC motor and vary its speed}

\newpage

\subsection{The program}
{ We modeled the Fuzzy logic controller established on matlab in a simple program on Arduino IDE.\\
We modeled the membership functions of the inputs: the error and the change of the error in mathematical equation (We already have (e) and (de) so we project on the curves of our membership functions to obtain the ordinate which varies between 0 and 1.
This ordinate will then be projected on the curves of the membership functions of our output (Kp,Ki,Kd).\\
The choice of the curves to consider will depend on condition If-Then which correspond to the rules established previously, table (2,3,4).
\\ \\
The code can be found on this link : https://github.com/aegniss/FUZZY-PID.git
}

\subsection{Experiment Results}
{ Once the code is launched, with a setpoint of 2000 rpm, we obtain the results shown in figure (19) ''the error is represented in blue and the output in green'':}
\FloatBarrier
\begin{figure}[htp]
    \centering
    \includegraphics[width=12cm]{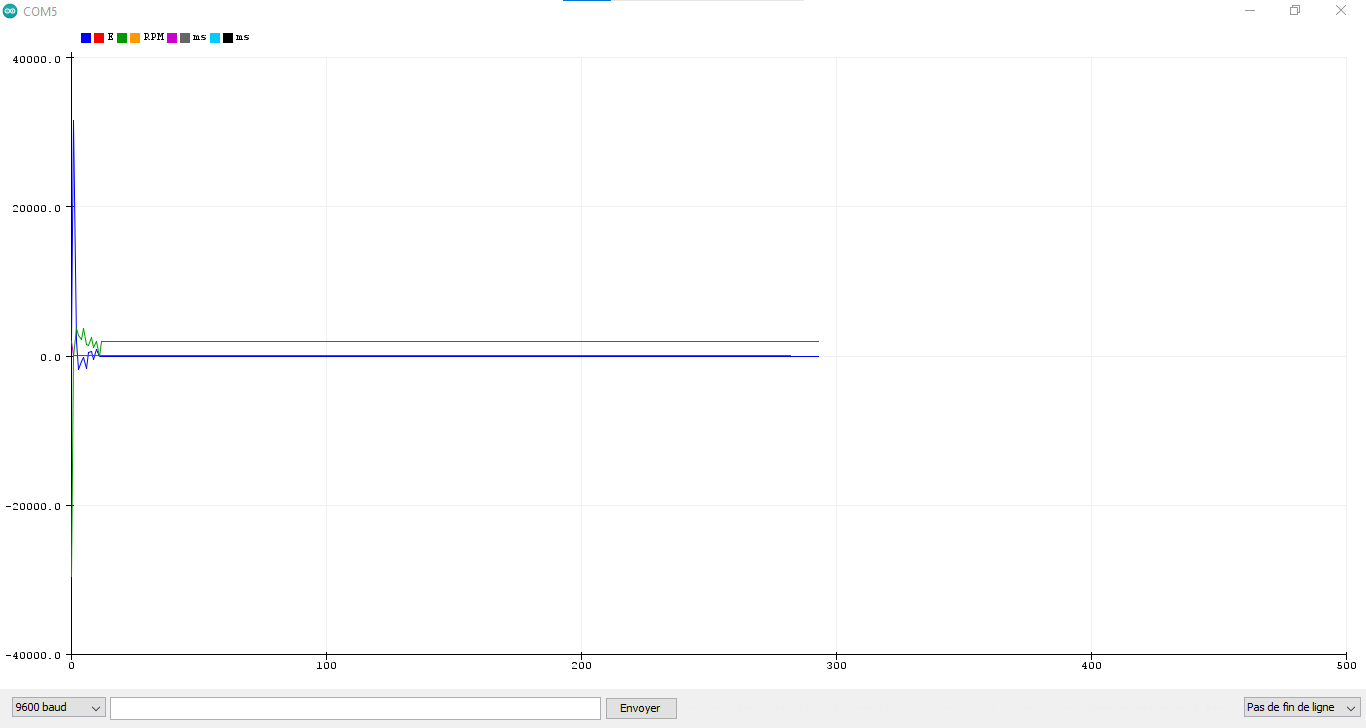}
    \caption{ Plot of the DC motor Error and Output in RPM}
    \label{fig: arduino plot}
     \includegraphics[width=6.5cm]{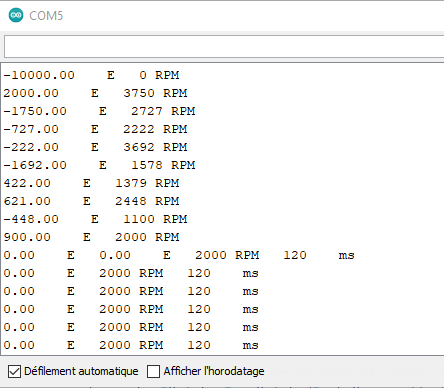}
    \caption{ Evolution of error and speed as a function of time}
    \label{fig: arduino plot}
\end{figure}
\FloatBarrier
{ In figure (19) and figure (20) we can see that the DC motor reaches the set point very quickly (at exactly 120ms), this is confirmed by the error trace that we can see in blue which is cancelled at the same time. We also notice that we have an overshoot before stabilization which is due to the high current absorbed at startup by the MCC. 
These observed results have a better response time than the fuzzy PID in theory but the latter presents a considerable overshoot at start point which is a negative point.
}

\section{Conclusion}
{
\textbf{Project objectives:}\\
In this project, we designed a DC motor whose speed can be controlled by a PID controller. The proportional, integral and derivative gains (KP, KI, KD) of the PID controller are adjusted using the fuzzy logic method. \\
The study shows that both the precise characters of PID controllers and the flexible characters of fuzzy controllers are present in the auto-tuning fuzzy PID controller.\\
The simulation results show that the designed self-tuning PID controller achieves the good dynamic behavior of the DC motor, perfect speed tracking with short rise and settling times, zero overshoot, steady state error null and thus gives better performance compared to the conventional PID controller.\\
}
{
\textbf{\\Difficulties encountered during the experiment :}\\
- Finding the right tuning method for the PID controller.\\ 
- Calculation of the derivative of the error in very small time intervals when coding on Arduino.\\
- Choice of the appropriate speed sensor given the very short measurement time and the speed of rotation of the DC motor (In our case the IR sensor was not precise enough during the measurement).\\
- Difficulty to implementing fuzzy logic on PIC microcontroller.\\
- Presence of a speed peak due to the amount of current absorbed at the start of the DC motor.\\
}
{
\textbf{\\Possible improvements to the project:}\\
- An alternative starting method to avoid speed peaks during start-up.\\
- Possibility to vary the speed (the setpoint) during the operation of the motor.\\
- Use of a more advanced microcontroller than the Arduino UNO.\\
- Possibility of entering the setpoint from a keyboard.\\
- Tuning of the PID parameters using genetic algorithms.\\
}

\newpage

\section{References}

{  [1]\; Ahmed Alhussain Ali Ahmed ; Samar Mohammed Abdullah ; Mohamed Omer Babiker   Mohamed Ali ; Abdulmajeed Abdulmonim Abdullatif Abdulrahman.\\}
{ \textit{ “DC Motor Control of Speed Using FUZZY-PID Controller”- A Project Submitted in Partial Fulfillment for the Requirements of the Degree of B.Sc. (Honor) In Electrical Engineering. }. \\}

{ [2]\; Guokun Xie et Kai Zheng and Yajuan Jia.\\}
{ \textit{"Design of Fuzzy PID Temperature Control System” }. \\}

{ [3]\;  K. Hachemi ; B. Mazari ; H. Oirkozek ; A. Al Jazi ; M. Laouer.\\}
{ \textit{ “Etude Comparative des Régulateurs PID et Flou : Autopilotage d'un Moteur Synchrone à Aimant Permanent ”  }.\\ }

{ [4]\; Umesh Kumar Bansal et Rakesh Narvey.\\}
{\textit{"Speed Control of DC Motor Using Fuzzy PID Controller”}.\\ }

{ [5]\; Alfatih Yassin Awad Babiker.\\}
{ \textit{“DC Motor Speed Control Using Fuzzy Logic and Proportional-Integral-Derivative Controller”-Thèse de Master d’ingénierie informatique et reseaux }. \\}

{ [6]\; F. Martin McNeill et Ellen Thro.\\ }
{ \textit{“THE FUZZY WORLD” }. \\}

{ [7]\; Antonio Visioli.\\}

{ \textit{“FUZZY LOGIC BASED TUNING OF PID CONTROLLERS FOR PLANTS WITH UNDER-DAMPED RESPONSE”  }.\\ }

{ [8]\; DEBASMITA PATTANAIK ; BONANI SAHU et DEVADUTTA SAMANTARAY.\\}
{ \textit{“MICROCONTROLLER BASED IMPLEMENTATION OF A FUZZY KNOWLEDGE BASED CONTROLLER”  }.\\ }

{ [9]\; \textit{“Modelling and Control of DC Motor”}. \\}

\end{document}